\documentstyle[12pt,titlepage,epsf]{article}
\begin{document}

\begin{center}
{\large\bf KINEMATIC DYNAMO WAVE IN THE VICINITY OF THE SOLAR POLES}
\end{center}

\begin{center}
V. M. Galitski$^1$ and  D. D. Sokoloff$^2$
\end{center}

\begin{center}
{\it $^1$ Theoretical Physics Institute, University
of Minnesota, 55455, USA
\newline $^2$ Department of Physics, Moscow State University,  119899, Russia
}
\end{center}

{\footnotesize\noindent We consider a dynamo wave in the solar
convective shell for the kinematic $\alpha\omega$-dynamo model. 
The spectrum and eigenfunctions of the corresponding equations are
derived analytically with the aid of the WKB method.  
Our main aim here is to investigate the dynamo wave behavior
in the vicinity of the solar poles. Explicit expressions
for the incident and reflected waves are obtained. The reflected wave
is shown to be relatively weak in comparison to the incident wave. 
The phase shifts and the ratio of amplitudes of the two waves are found.}


\section{Introduction}

The solar cycle is a well-known manifestation of the magnetic activity
of the Sun. The physical mechanism responsible for this activity is
thought to be the dynamo generation of the large-scale solar magnetic field
(see {\em e.g.} Zel'dovich, Ruzmaikin and Sokoloff, 1983).  There are
a number of numerical models of the solar and stellar dynamo waves
which reproduce many features of the observed solar magnetic
activity (Brandenburg, 1994; R\"udiger and Brandenburg, 1995; Tobias,
1997 etc.).

It is important to support these numerical investigations by
approximate analytical solutions of the dynamo equations.  Recently,
new analytical methods of solving the mean field dynamo
equations have been developed for the case of a very intensive generation.
However, analytical solutions were obtained only in the
simplest cases. Kuzanyan and Sokoloff (1995) derived the asymptotic
solution of the Parker migratory dynamo equations (Parker,
1955). Meunier {\it et al.} (1997) and Bassom {\it et al.} (1997) have
considered asymptotic solutions of the Parker migratory dynamo in
the nonlinear regime.

These results, as well as the Parker's equations themselves, have a
limited domain of applicability. In particular, they do not reflect
the details of the dynamo wave behavior in the very vicinity of the
poles. In the present paper, we consider dynamo effects in the
framework of a more general model, taking into account the convective
shell curvature. In the limiting case of very large dynamo numbers, we
derive consistently the spectrum and eigenfunctions of the dynamo
equations and describe the dynamo wave near the solar pole.

According to the modern observational results, the dynamo wave and the
wave of sunspots, observed in the butterfly-diagrams, propagate
equator-wards within the main spatial domain ({\em i.e.} far from the
solar equator and solar poles).  Besides, there is a weak dynamo wave
in the vicinity of the solar poles which was observed by Makarov and
Sivaraman (1983).  This wave propagates pole-wards. In this paper, we
explore this polar dynamo wave by solving the generalized equations of
the Parker migratory dynamo. We show that the incident dynamo wave
reflects from the pole. The reflected wave is relatively weak
compared to the incident wave.  We demonstrate also that the process
of dynamo wave reflection reduces to a phase jump and the reflection
is not accompanied by any sharp changes of the wave amplitude.

Our paper is structured as follows: In Sec. 2, we derive the equations
in the Parker's model from the general equations of the mean field
electrodynamics. In Sec. 3, we solve these equations asymptotically
with the aid of the WKB method, using the fact that the dynamo number
is very large for the Sun.  We obtain the spectrum and the
eigenfunctions describing the dynamo wave far from the pole. In Sec.
4, we reduce the equations to a simplified form valid in the vicinity
of a pole.  In Sec. 5, we solve these asymptotic equations and show
that our solution describes two waves: the incident wave and the reflected
wave.  In Sec. 6, we match two asymptotics found in Secs. 3 and 5 and
derive the ratio of the amplitudes of the incident and reflected waves
and the corresponding phase shifts.  In summary section, we discuss
our results and make some simple numerical estimates.

\section{Basic Equations}

Here we obtain a generalization of the  well-known Parker
migratory dynamo equations which takes into account the curvature of
the convective shall.  The large-scale magnetic field generation in a
turbulent flow of a differentially rotating electrically
conducting fluid is governed by the following equation (see Krause
and R\"adler, 1980):

\begin{equation}
\label{MFE}
{\partial \bf B \over \partial t} = {\bf \nabla} {\bf \times} \left( \alpha {\bf B} \right)+
{\bf \nabla \bf \times} \left( {\bf v \bf\times \bf B} \right) + \beta \Delta {\bf B},
\end{equation}
where $\bf B$ and $\bf v$ are the large-scale (mean) magnetic and
velocity fields correspondingly, $\alpha$ is the helicity
coefficient and $\beta$ is the turbulent diffusivity.

Looking for a kinematic axisymmetric eigensolution of
Eq.(\ref{MFE}), we present the magnetic field as a
superposition of the poloidal and toroidal fields as follows:
\begin{equation}
\label{torpol}
{\bf B} ({\bf r},t) =
\left[ {\bf B}_p({\bf r}) + 
{\bf B}_t ({\bf r}) \right] e^{\gamma t},
\end{equation}
where ${\bf B}_t = \left( 0,\, 0,\, \tilde B \right)$ is the azimuthal
component of the magnetic field, 
${{\bf B}_p} = R\, {\rm rot} \, \left( 0,\,
0,\, \tilde A \right)$ is the radial component of the magnetic field 
($R$ is the solar radius) and $\gamma$ is the eigenvalue to be found.
Thus, ${\rm Re \,} \gamma$ is the
magnetic field growth rate and $2 \pi ({\rm Im \,} \gamma)^{-1}$ is
the dynamo wave period. 

In the new terms, Eq.(\ref{MFE}) takes on the following
dimensionless form:
\begin{equation}
\label{dimA}
\gamma\tilde A ={R_{\alpha}}\alpha \tilde B+
{1 \over r} {{\partial ^2} \over {\partial r^2}}({\tilde A}r)+ {1
\over {r^2}}
{\partial \over {\partial \theta}}
\left[ {1 \over {{\rm sin} \theta}}{\partial \over {\partial \theta}}
({\tilde A} {\rm sin} \theta) \right],
\end{equation}
$$\gamma \tilde B = {R_{\omega}} G{\partial \over{\partial
\theta}}({\tilde A}r{\rm sin} \theta)
+ {R'_{\omega}}G'{\partial \over {\partial r}}({\tilde A}r {\rm
sin} \theta) - {R_{\alpha}}{1 \over r}{\partial\over {\partial
r}}
\left[ \alpha{\partial \over {\partial r}} ({\tilde A}r) \right] -
$$
\begin{equation}
\label{dimB}
- {R_{\alpha}}{1 \over r^2}{\partial \over {\partial \theta}}
\left[ {\alpha \over {{\rm sin}\theta}}
{\partial \over {\partial \theta}}({\tilde A}{\rm sin} \theta)
\right]
+ {1\over r}{{\partial ^2} \over {\partial r^2}}(\tilde B r) + {1 \over
{r^2}}{\partial \over {\partial \theta}}
\left[ {1\over{{\rm sin}\theta}}
{\partial\over {\partial \theta}} (\tilde B {\rm sin} \theta ) \right],
\end{equation}
where we have introduced the following dimensionless constants:

$$ R_{\alpha}={{\alpha_{max} R} \over \beta},\,\,
R_{\omega}={R^2 \over \beta}G_{max},\,\, R'_{\omega}={R^2 \over
\beta}G'_{max}.
$$ 
Here, $G = { \displaystyle 1 \over \displaystyle r}
{\displaystyle \partial\Omega \over \displaystyle\partial r}$
is the radial gradient of the mean angular velocity and
$G'={\displaystyle 1 \over \displaystyle r}
{{\displaystyle\partial {\mit \Omega}} \over
\displaystyle{\partial\theta}}$ is its meridional gradient.  Note, that 
lengthes are measured in units of the solar radius, time in units of
the diffusion time $\tau_{\rm dif} = {R^2 / \beta}$, $\alpha$
and $G$ in units of their maximal values.

We suppose that the differential rotation is more intensive than
the mean helicity and that $G(r,\theta)$ weakly depends on the
latitude.  Thus, the terms in Eq.(\ref{eqB}) which contain
$R_{\alpha}$ and $G'$ can be omitted. This approximation is known as
$\alpha\omega$-dynamo model.

We also suppose that the process of intensive magnetic field
generation takes place in a thin shell . Parker (1955) has noted, that
in this case, two-dimensional dynamo equations could be reduced to
effectively one-dimensional equations by their averaging over $r$ (see
also Kuzanyan and Sokoloff, 1996). Below, we explore the
corresponding one-dimensional problem.

All the assumptions, made above, yield a
significant simplification of Eqs.(\ref{dimA}, \ref{dimB}) and we have:
\begin{equation}
\gamma A = \alpha(\theta) B + {d \over {d
\theta}} \left[ {1 \over {\sin \theta}}
{d \over {d \theta}} (A
\sin \theta) \right],
\label{eqA}
\end{equation}
\begin{equation}
\gamma B =D G(\theta) {d \over {d
\theta}} (A \sin \theta) + {d \over {d \theta}} \left[ {1 \over
{\sin \theta}} {d \over {d \theta}} (B \sin
\theta) \right].
\label{eqB}
\end{equation}
Here $D=R_{\alpha}R_{\omega}$ is the dimensionless dynamo number,
which characterizes the intensity of the generation sources, function
$A(\theta)= R_{\alpha}^{-1} \langle \tilde A(\theta, r) \rangle$ is
proportional to the averaged azimuthal component of the
vector-potential and $B(\theta) = \langle \tilde B(\theta,r) \rangle$
is the averaged azimuthal component of the magnetic field ( $\langle
...  \rangle$ means averaging over $r$), $\theta$ is the latitude
measured from the solar pole.  Helicity coefficient $\alpha(\theta)$ is also
averaged over the shell section.  We suppose that $\alpha(0)
\ne 0$.  Note, that after averaging, the diffusive terms take the following
form (see also Proctor \& Spiegel, 1991):
$$\left\langle {1 \over r} 
{\partial^2 \over \partial^2 r} \left(\tilde B r \right) \right\rangle = -\mu^2
B,\,\,\,\, 
\left\langle {1 \over r} {\partial^2 \over \partial^2 r} \left( \tilde A r
\right) \right\rangle = - \mu^2 A$$
and they can be taken into account by a redefinition of
eigenvalue $\gamma$.

Mention, that a typical estimate of the dynamo-number for the Sun is
$|D| \approx - 10^4$ or even more. Below, we investigate
Eqs.(\ref{eqA}, \ref{eqB}) analytically taking into account the large
value of the dynamo number. We derive the asymptotical spectrum
corresponding to these equations and the asymptotic behavior of the
dynamo wave. We consider two spatial domains: The first
one is the main domain which is arranged far both from the poles and
from the equator and the second one is the domain near a solar pole
($\theta \ll 1$).

Let us note that the simplest approximate form of the one-dimensional
dynamo equations, was obtained  phenomenologically by Parker in 1955
(see also Stix, 1989). Those equations, known as Parker's equations
follow from equation system (\ref{eqA}, \ref{eqB}). They
appear as a first approximation for the case of very short dynamo waves:
\begin{equation}
\label{Parker_A}
{\gamma A} = \alpha(\theta) B +{d^2 A
\over d\theta^2};
\end{equation}
\begin{equation}
\label{Parker_B}
{\gamma B} = D G(\theta) \sin{\theta}
{dA \over d\theta} + {d^2 B \over
d\theta^2}.
\end{equation}
Let us mention that using the Parker's equations one can obtain the
correct expressions for eigenvalues $\gamma$ in the leading
approximation.  However, the eigenfunctions do not coincide, even in
the main approximation, with the corresponding solution of
Eqs.(\ref{eqA},\ref{eqB}).  Note also that equations similar to
(\ref{Parker_A}, \ref{Parker_B}) were solved numerically
in the nonlinear regime by Jennings (1991) in the main domain.

\section{Asymptotic solution in the main domain}
To obtain the asymptotic solution of Eqs.(\ref{eqA}, \ref{eqB}) in the
main domain, we use the WKB approach (see {\em e.g.} Maslov and
Fedorjuk, 1981).  Si\-mi\-lar method was ap\-plied by Kuza\-nyan and
Sokoloff (1995) to solve the Par\-ker's equ\-a\-tions.  In this
section, we follow their treatment to obtain the solution of the more
general equations under discussion.

Let us rewrite Eqs.(\ref{eqA}, \ref{eqB}) in the spectral form
explicitely:
\begin{equation}
\label{Sp_form}
\hat H {A(\theta) \choose B(\theta)}=
\gamma {A(\theta) \choose B(\theta)},
\end{equation}
where $\hat H$ is a linear differential operator:
\begin{equation}
\label{H}
\hat H =
\left(
\begin{array}{cc}
\frac{\displaystyle d}{\displaystyle d\theta} 
{\displaystyle 1 \over \displaystyle\sin{\theta} }
{\displaystyle d \over {\displaystyle d\theta} } \sin{\theta}&
\alpha(\theta)\\
\,\, & \,\,\\
D G(\theta) 
{\displaystyle d \over {\displaystyle d\theta}} \sin{\theta}&
{\displaystyle d \over {\displaystyle d\theta} } 
{\displaystyle 1 \over \displaystyle\sin{\theta} }
{\displaystyle d \over {\displaystyle d\theta} } \sin{\theta}
\end{array}
\right).
\end{equation}

Let us present the eigenvector and eigenvalues in the following form:
\begin{equation}
\label{as}
{A \choose {\varepsilon^2 B} } =
\left[
{\mu \choose \nu} + \varepsilon{\mu_1 \choose \nu_1} + \ldots
\right]
{\rm exp} \left[{{i S} \over {\varepsilon}}\right],
\,\,\, {\rm where}\,\,\, \varepsilon = |D|^{-(1 / 3)},
\end{equation}
\begin{equation}
\label{gamma_as}
\gamma = {1 \over \varepsilon^2} \Gamma_0 +
{1 \over \varepsilon} \Gamma_1 + \ldots.
\end{equation}
Here $\mu$, $\mu_1$, $\nu$, $\nu_1$ and $S$ are complex
functions of the latitude and $\Gamma_0$, $\Gamma_1$ are complex
constants.  Emphasize that value $\varepsilon = |D|^{-{1 \over 3}}$ is
the true parameter of our asymptotic expansion and its smallness
is the main condition of applicability of our solution.

To find the spectrum of Eq.(\ref{Sp_form}), we have to formulate 
boundary conditions. Worledge {\it et al.} (1997) have emphasized the
role of these conditions. They showed that the solution in the linear regime
is very sensitive to the changes in the boundary conditions  
(see Tobias {\it et al.}, 1997 for the nonlinear regime).  
Our asymptotic WKB expansion is not applicable near the boundary, i.e.
in the very vicinity of the poles and the solar equator. Thus, we can
not formulate boundary conditions in the explicit form in the framework
of the asymptotic theory.  However, we can match the asymptotic WKB
solution applicable in the main domain with the solution applicable
near the pole. The boundary condition for the latter is that the
magnetic field is limited everywhere including the pole. To perform
the matching one must require that the asymptotic solution decays for
$\theta \to 0$ and $\theta \to {\pi \over 2}$. Otherwise, the field
grows exponentially in the vicinity of the boundary and the matching
is impossible. The other requirement is that the asymptotic solution is a
smooth function. These two conditions are found to be sufficient to
obtain an asymptotic spectrum.

Note, that the dynamo number can have any sign.  The sign determines
the direction of the dynamo wave propagation. We choose the most
realistic case in our solution. Using the observational fact that the
dynamo wave propagates equator-wards in the main domain, we accept the
negative sign for $D$ that corresponds to this behavior.

Equations of the first approximation can be obtained by
substituting Eqs.(\ref{as}) and (\ref{gamma_as}) into
Eq.(\ref{Sp_form}) and equating the terms of the minimal power of
$\varepsilon$. This yields:
\begin{equation}
\label{1a}
\left(
\begin{array}{cc}
\Gamma_0 + k(\theta)^2&
- \alpha (\theta)\\ i G(\theta) \sin{\theta} k(\theta)&
\Gamma_0 + k(\theta)^2
\end{array}
\right)
{\mu(\theta) \choose \nu(\theta)} = 0.
\end{equation}
A non-trivial solution of this equation exists only if the determinant
of the matrix in the left-hand side of Eq.(\ref{1a}) vanishes. This
condition leads to so-called Hamilton-Jakobi equation (we use the
standard terminology of the WKB-method, see e.g. Landau and Lifshitz,
1958):
\begin{equation}
\label{HJ}
\left( \Gamma_0 + k^2 \right)^2 + i \hat\alpha k = 0,
\end{equation}
where $k(\theta)=\frac{\displaystyle d S}{\displaystyle d
\theta}$ is the ge\-ne\-ra\-li\-zed
mo\-men\-tum (by ana\-lo\-gy with the semi\-clas\-sical appro\-xi\-ma\-tion in
quan\-tum mecha\-nics). We have also introduced a new function
$\hat\alpha(\theta) = G(\theta) \alpha(\theta) \sin{\theta}$.

Note that the Hamilton-Jakobi equation in our problem coincides with
the one for the Parker's equations. Its solution is described in the
paper of Kuzanyan and Sokoloff (1995).  Here, we will recall shortly
the reasonings that allow to calculate spectral parameter $\Gamma_0$
from the Hamilton-Jakobi equation.

The accepted decay condition (the solution 
should be small near the boundary) can be  rewritten 
in the terms of the generalized momenta as follows:
$$
\left. {\rm Im \, } k \right|_{\theta \to {\pi \over 2}} > 0, \,\,\,
\left. {\rm Im \,} k \right|_{\theta \to 0} < 0.
$$ 
These expressions play the role of boundary conditions.

\begin{figure}
\epsfxsize=7.5cm
\centerline{\epsfbox{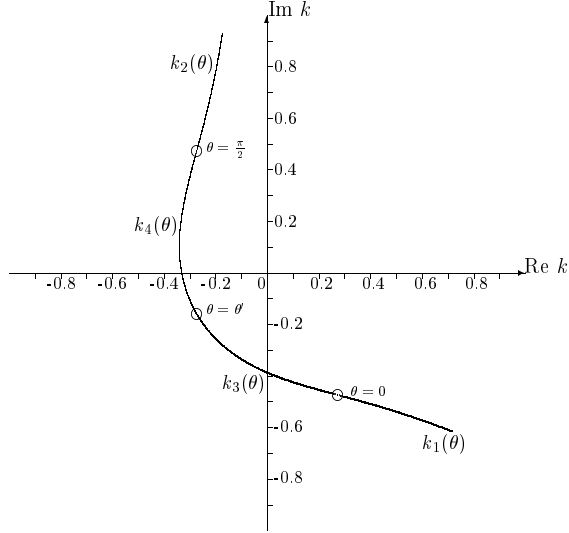}}
\caption{\rm\protect\footnotesize 
Roots of Hamilton-Jakobi equation
(\protect\ref{HJ}) in
the complex plane $k$ plotted as functions of latitude $\theta$ for the
chosen value of $\Gamma_0$. The points of matching of different
branches are circled and the corresponding values of $\theta$ are given.}
\end{figure}

Equation (\ref{HJ}) is the algebraic equation of the fourth order
and thus it possesses four branches of roots. However, none of
these branches satisfies the boundary conditions. It means that
we should construct our solution by matching two (or more)
branches. To explore this case, we denote 
$$ H(k,\theta)=\left( \Gamma_0 + k^2
\right)^2 + i k \hat\alpha(\theta).
$$ 
In  terms of function $H(k,\theta)$, it is easy to formulate the
conditions of crossing of two branches at a point $\theta'$:
\begin{equation}
\label{cross}
\left\{
\begin{array}{l}
H[k(\theta'),\theta']=0;\\
\,\,\\ 
{{\displaystyle\partial H} \over {\displaystyle\partial k} }
[k(\theta'),\theta'] =0.
\end{array}
\right.
\end{equation}
These equations can be solved explicitly and we have
$$
\Gamma^{(1,2)}_0 = \frac {3 \hat\alpha(\theta')^{2 \over 3}
e^{\pm i {\pi \over 3}}} {2^{8 \over 3}}, \,\,\,
\Gamma^{(3)}_0 = -\frac {3 \hat\alpha(\theta')^{2 \over 3}}{2^{8 \over 3}}.
$$


As one can see, the real part of $\Gamma_0^{(3)}$ is negative, which
corresponds to a decaying solution. The two other conjugated values
correspond to growing solutions of our interest. The leading mode is
the most important from the physical point of view. Thus, we choose
$\theta'$ to be the point where function $\hat\alpha(\theta)$ reaches
its maximum (see Fig.1). It was shown (Galitski and Sokoloff, 1998),
that only this choice of $\Gamma$ leads to smooth eigenfunctions of
the Parker equations.

From Eq.(\ref{1a}), we have:
\begin{equation}
\label{munu}
 {\mu(\theta) \choose \nu(\theta)} = {{\Gamma_0 + k^2} \choose
 {-i k \sin{\theta}}} \sigma(\theta),
\end{equation}
where $\sigma(\theta)$ is to be defined from the equation of
the second approximation. To obtain this equation, we
substitute Eqs.(\ref{as}) and (\ref{gamma_as}) into Eq.(\ref{Sp_form})
and equate the terms of the first power of $\varepsilon$. This yields:
\begin{equation}
\label{2a}
\left(
\begin{array}{cc}
\Gamma_0 + k^2 & - \alpha\\
-i k \sin{\theta}&\Gamma_0 + k^2
\end{array}
\right)
{\mu_1 \choose \nu_1} = {{ \left[ ik'+ik {\, \rm ctg \,}{\theta}
-\Gamma_1 \right] \mu + 2 i k \mu'}
\choose
{\left[ ik'+ik {\, \rm ctg \,}{\theta} -\Gamma_1 \right] \nu + 2 i k \nu'-
\left[ \mu \sin{\theta} \right]'}}.
\end{equation}
Here, for the sake of simplicity we neglect the term containing 
$G'(\theta)$, supposing that
the angular rotation weakly depends on the latitude in the main
domain.  Note, that (\ref{2a}) differs from the
corresponding expression for the Parker's equations.


We see that the matrixes in the left-hand sides of Eqs.(\ref{1a}) and
(\ref{2a}) are identical.  As we have seen the corresponding matrix is
degenerate.  Thus, Eq. (\ref{2a}) possesses solutions only if the
Fredholm resolvability condition is satisfied, {\em i.e.} if the vector in
the right-hand side of Eq.(\ref{2a}) is ortoganal to the eigenvector of
the adjoint equation. It yields so-called transport equation:
$$
\left[
  \Gamma_1 - i k' \left( 1 + \frac{2 k^2}{\Gamma_0 +k^2} \right) - {\,
    \rm ctg \,}{\theta} \left( 2 i k - \frac{\hat\alpha}{2
      (\Gamma_0+k^2)} \right) \right] \sigma(\theta) =
$$
\begin{equation}
\label{eqsig}
= \left[ 2 i k - \frac{\hat\alpha}{2 (\Gamma_0+k^2)}
\right] \sigma(\theta').
\end{equation}
Using the fact that the function in the brackets in the
right-hand side of equation (\ref{eqsig}) vanishes at $\theta'$
(recall that $\theta'$ is the point at which $\hat\alpha$ is
maximal) and following the standard procedure of the WKB theory (see
Maslov and Fedorjuk, 1981) we can find the spectrum.
First, let us denote the following functions:
\begin{equation}
\label{P}
P(\theta)=2 i k - \frac{\hat\alpha}{2 (\Gamma_0+k^2)},
\end{equation}
and
\begin{equation}
\label{Q}
Q(\theta) = i k' \left( 1 + \frac{2 k^2}{\Gamma_0 +k^2} \right)
+ P(\theta) {\rm {\, \rm ctg \,}}{\theta}.
\end{equation}
We suppose that $P(\theta), \, Q(\theta), \, \sigma(\theta)$ are
analytical functions in the vicinity of $\theta'$.
Furthermore, we present all these functions in the form of
Taylor series:
\begin{equation}
\label{PT}
P(\theta) = \left[ P_0 + P_1(\theta) (\theta - \theta') \right]
(\theta - \theta'),
\,\, {\rm where} \,\, P_0= 3 i k'(\theta');
\end{equation}
\begin{equation}
\label{QT}
Q(\theta) =Q_0 + Q_1(\theta) \left( \theta-\theta' \right),
\,\, {\rm where} \,\, Q_0= {3 i k'(\theta') \over 2}.
\end{equation}
For $\sigma_n$ we have
\begin{equation}
\label{sigmaT}
\sigma_n(\theta) = \left( \theta -\theta'\right)^n
\left[ C_n + C_{n+1} \left( \theta -\theta' \right) \right],
\end{equation}
where $n$ is an integer parameter, which classifies
the eigenvalues.  Substituting (\ref{PT}),(\ref{QT}) and
(\ref{sigmaT}) into (\ref{eqsig}) and setting $\theta=\theta'$, we have
\begin{equation}
\label{prespectr}
P_0 n = \Gamma_{1,n} - Q_0.
\end{equation}
From here we can easily obtain the spectrum:
\begin{equation}
\label{spectre}
\Gamma_{1,n} = 3 i k'(\theta') \left( n + {1 \over 2} \right),
\,\,\, n=0,1,2, \ldots.
\end{equation}

The explicit expression for $\sigma$ is
\begin{equation}
\label{sigma}
\sigma(\theta) = {1 \over \sin{\theta}} {\rm exp} \left\{
\int \frac{\Gamma_1 - i k' \left( 1 + 
\frac{\displaystyle 2 k^2}{\displaystyle \Gamma_0 +k^2} \right)}
{2 i k - \frac{\displaystyle \hat\alpha}
{\displaystyle 2 (\Gamma_0+k^2)}} {d \theta}
\right\} .
\end{equation}
We emphasize that Eqs.(\ref{spectre}) and (\ref{sigma}), in contrast to
the Hamilton-Jakobi equation, differ from the corresponding
expressions obtained by Kuzanyan and Sokoloff (1995).  Surprisingly,
our solution has a simpler structure despite the fact that the initial
equations in our problem are more complicated than the Parker equations.

\section{Equations Near The Pole}

We now proceed to investigate equations (\ref{eqA}, \ref{eqB}) in the
second domain near the solar pole. In the following, we have two small
values: $\theta$ and $\varepsilon$.  Since $\theta \ll 1$, we can rewrite equation
Eq.(\ref{eqA}) in the following form:
\begin{equation}
\label{eqA1}
\gamma A = \alpha(0) B + {{d^2A} \over {d\theta^2}}+
{1 \over \theta}{ {d A} \over {d \theta}} - {1 \over {\theta^2}}
A.
\end{equation}
Here we keep only the terms of the minimal power of $\theta$.

Taking into account results of Sec. 3, it is reasonable to assume that
$|A| \sim \varepsilon^2 |B|$ and each differentiation of
the field multiplies it by $\varepsilon^{-1}$ (formally we can
write ${d \over {d \theta}} \sim {1 \over \varepsilon}$).  Our
assumptions will be confirmed by the results obtained below.
This yields the following estimates of the terms in (\ref{eqB}):

$$B'' \sim {1 \over \varepsilon^4} |A|, \,\,
\gamma B \sim {1 \over \varepsilon^4}|A|, \,\,
{\rm {\, \rm ctg \,}}{\theta} B'
\sim {1 \over \theta\varepsilon^3} |A|, \,\,
{1 \over \sin{\theta}^2} B \sim {1 \over {\theta^2
\varepsilon^2}} |A|, \,\,$$ 
$$D G \sin{\theta} A' \sim {\theta \over
\varepsilon^4} |A|, \,\,
D G \cos{\theta} A \sim {1 \over \varepsilon^3} |A|.$$

The terms corresponding to the last two
estimates can be neglected and  Eq.(\ref{eqB}) reads:
\begin{equation}
\label{eqB1}
\gamma B=  {{d^2B} \over {d\theta^2}}+
{1 \over \theta}{ {d B} \over {d \theta}} - {1 \over {\theta^2}}
B.
\end{equation}

Equations (\ref{eqA1}, \ref{eqB1}) can be rewritten in the 
following compact form:
\begin{equation}
\label{JA}
\hat J_1(x) A(x) = {\alpha(0) \over \gamma} B(x);
\end{equation}
\begin{equation}
\label{JB}
\hat J_1(x) B(x) = 0,
\end{equation}
where we have introduced a new variable $x = \sqrt{-\gamma}
\theta$ and
a differential operator 
$$
\hat J_1 (x) = {d^2 \over dx^2} + {1
\over x}{d \over dx} +
\left( 1 - {1 \over x^2} \right).
$$ 
Let us note that in Eqs.(\ref{JA}, \ref{JB})
$\gamma$ plays the role of an external parameter. Its value was
obtained in Sec. 3.

\section{Dynamo Wave in The Vicinity of The Solar Pole.}
In this section, we solve equations Eqs.(\ref{JA}, \ref{JB}) obtained above.
Equation (\ref{JB}) is the homogeneous Bessel's equation of the first
order for $B$ and Eq.(\ref{JA}) is the inhomogeneous Bessel's equation
of the first order for $A$. The magnetic field must be finite
for all $x$. It means that we should take Bessel
function of the first order $J_1(x)$ as a solution of
Eq.(\ref{JB}).  Hence,
\begin{equation}
\label{B(x)}
B(x) = 2 C_B J_1(x) = C_B \left[ H_1^{(1)}(x) + H_1^{(2)}(x)
\right].
\end{equation}
Here $C_B$ is a constant and $H_1^{(1,2)}(x)$ are Hankel
functions of the first type and the second type of the first order. 
Note, that Hankel functions diverge at the pole ($\theta=0$),
but their sum is finite. As we shall see below, $H_1^{(1)}(x)$ describes the
incident wave and $H_1^{(2)}(x)$ describes the reflected wave.

The general solution of Eq.(\ref{JA}) is a sum of general
solution of the corresponding homogeneous equation and a
particular solution of the inhomogeneous equation. To find 
particular solution $A_1(x)$, note that the Wro\'nski
determinant of Hankel functions is
\begin{equation}
\label{Wron}
H_1^{(1)}(x) H_1^{(2)}(x)' - H_1^{(1)}(x)' H_1^{(2)}(x) = - {4 i
\over \pi x},
\end{equation}
hence,
$$
A_1(x)
= - {\pi\alpha(0) \over 4 i \gamma} C_B
\left[
H_1^{(2)}(x)
\int \left[ H_1^{(1)}(x)^2 + H_1^{(1)}(x)H_1^{(2)}(x)\right] x dx
\right. -
$$
\begin{equation}
\label{parsol}
\left.
- H_1^{(1)}(x)
\int \left[ H_1^{(2)}(x)^2 + H_1^{(1)}(x)H_1^{(2)}(x)\right] x dx
\right].
\end{equation}

Summarizing, we have
\begin{equation}
\label{fin}
\left\{
\begin{array}{l}
A(x) = C_A \left[ H_1^{(1)}(x) + H_1^{(2)}(x) \right] + A_1(x),\\
B(x) =C_B \left[ H_1^{(1)}(x) + H_1^{(2)}(x) \right].
\end{array}
\right.
\end{equation}
Two complex constants $C_A$ and $C_B$ in (\ref{fin}) are not
independent.  Really, fields $A(x)$ and $B(x)$ represent two
different components of a true eigenvector for Eq.(\ref{eqA}, \ref{eqB}).
The norm of this eigenvector is an arbitrary quantity, but its
orientation is prescribed.  Our asymptotic solution
(\ref{fin}) must satisfy this condition too.  To find a
connection between $C_A$ and $C_B$, one should consider a spatial
domain where both asymptotics (\ref{as}) and (\ref{fin}) are
valid.  This domain is characterized by the following condition:
\begin{equation}
\label{zone}
\varepsilon \ll \theta \ll 1
\end{equation}

Since $\left| x \right| \gg 1$, we can use the well-known asymptotic form
of Hankel functions for large arguments:
\begin{equation}
\label{asH1}
H_1^{(1)}(x) \approx \sqrt {2 \over \pi x} e^{i \left( x -{3 \pi
\over 4} \right)},
\end{equation}
\begin{equation}
\label{asH2}
H_1^{(2)}(x) \approx \sqrt {2 \over \pi x} e^{-i \left( x -{3 \pi
\over 4} \right)},
\end{equation}
i.e.
\begin{equation}
\label{asBxt}
B(x,t) \approx C_B \sqrt {2 \over \pi x}
\left[ e^{i \left( x -{3  \pi \over 4} \right)} +
e^{-i \left( x -{3 \pi \over 4} \right)} \right] e^{\gamma t}.
\end{equation}
Let us emphasize, that  constraint (\ref{zone})
is not a condition of applicability of Eq.(\ref{fin}), but a
condition of applicability of asymptotics (\ref{asH1}, \ref{asH2})
and (\ref{h}) (see Sec. 6).
These approximate presentations allow us to perform the matching in the
explicit form. However, the condition (\ref{zone}) can be satisfied only if 
$\varepsilon$ is extremely small. Otherwise (if it is not possible to find a 
domain where the value $\theta$ is both much lesser than unity and 
much greater than $\varepsilon$), one should perform the matching using
the general form of our asymptotic solution.    

Recalling that $x = \sqrt{- \gamma} \theta$, one can define the
phase surface as $$
\phi(\theta,t) = \pm {{\rm Re \,}\sqrt{-\Gamma_0} \over \varepsilon}\theta +
{{\rm Im \, }\Gamma_0 \over \varepsilon^2} t = {\rm const}.  $$ The
phase velocity is
\begin{equation}
\label{vph}
v_{\rm phase}=
\dot\theta
= \mp {1 \over \varepsilon}
\frac {{\rm Im \,} \Gamma_0}{{\rm Re \,}\sqrt{-\Gamma_0}}.
\end{equation}
If $v_{\rm phase}<0$ than the wave propagates pole-wards, if
$v_{\rm phase}>0$ than the wave propagates equator-wards.

Evaluating $v_{\rm phase}$ explicitly, we readily conclude that $H_1^{(1)}(x)$
corresponds to the incident wave and $H_1^{(2)}(x)$ corresponds to
the reflected wave.

The asymptotic form of $A(x,t)$ is $$ A(x,t) = \left[ \sqrt{2
\over \pi x} C_A + {\alpha(0) \over \gamma}
\sqrt{x \over 2 \pi} C_B \right]
e^{i \left(x - {3 \over \pi 4} \right) + \gamma t}+ $$
\begin{equation}
\label{asAxt}
+ \left[ \sqrt{2 \over \pi x} C_A - {\alpha(0) \over \gamma}
\sqrt{x \over 2 \pi} C_B \right]
e^{-i \left(x - {3 \over \pi 4} \right) + \gamma t}.
\end{equation}

\section{Matching of Asymptotics}

One can see from (\ref{vph}) that branches $k_3$ and $k_4$ (see Fig.1)
correspond to the incident wave. Now, we have to find a branch
of roots $k_i(\theta)$ that corresponds to the reflected
wave. We expect that the reflected
wave, if exists, should be very weak. It means that this wave 
decays going equator-wards. Thus, we must choose
branch $k_2$ to represent the reflected wave. Only this branch is
arranged in the upper part of the complex plane $k$ and
so it describes a decaying wave.

Now, we should match a linear combination of asymptotics
(\ref{as}) for the incident and reflected waves and asymptotics
(\ref{asAxt}, \ref{asBxt}).  For this aim, it is necessary to
evaluate approximate expressions for $k_{\pm}(\theta)$,
$\sigma_{\pm}(\theta)$, $A_{\pm}(\theta)$ and $B_{\pm}(\theta)$
for the case $\theta \ll 1$ (here and further +'s correspond to
the incident wave and -'s correspond to the reflected wave).
From equations (\ref{as}), (\ref{HJ}), (\ref{sigma}) and
(\ref{spectre}) after a rather long but straightforward algebra 
we have following expressions (for $\theta \to 0$):

\begin{equation}
\label{k+}
k_{+}(\theta) = k_0 -
\kappa e^{-i{\pi \over 12}}\sqrt{\theta},
\end{equation}
\begin{equation}
\label{k-}
k_{-}(\theta) = -k_0+
\kappa e^{5 i \pi \over 12}\sqrt{\theta},
\end{equation}
where $\kappa$ and $k_0$
\begin{equation}
\label{kap}
\kappa =
{ {\sqrt{\hat\alpha'(0)} \over {2^{1 \over 3} 3^{1 \over 4}
\hat\alpha(\theta')^{1 \over 6}}}} ,\,\,
k_0= {\sqrt{3} \hat\alpha(\theta')^{\frac{1}{3}} \over
2^{\frac{4}{3}}} e^{-i \frac{\pi}{3}},
\end{equation}
\begin{equation}
\label{sigpm}
\sigma_{\pm}(\theta) =
{1 \over \theta^{5 \over 4}} e^{\pm f_{\pm} \sqrt{\theta}},
\end{equation}
where $f_{\pm}$
\begin{equation}
\label{f}
f_{\pm} = {7 \over 2^{7 \over 2} 3^{3 \over 4}}
\sqrt{\hat\alpha'(0) \over \hat\alpha(\theta')} \left( 1 \pm i \right).
\end{equation}
Using these expressions, one can obtain the approximate form of
the solution of (\ref{eqA}, \ref{eqB}) in the main
domain:
\begin{equation}
\label{h}
{A \choose B}= C_{+} {i \rho \theta^{-{3 \over 4}} \choose
\beta \varepsilon^{-2} \theta^{-{1 \over 4}}} e^{ix} -
C_{-} {\rho \theta^{-{3 \over 4}} \choose
\beta \varepsilon^{-2} \theta^{-{1 \over 4}}} e^{-ix},
\end{equation}
where

$$
\rho=
\frac{3^{1 \over 4} \alpha(\theta')^{1 \over 6} \sqrt{\alpha'(0)}}
{2^{2 \over 3}} e^{i \pi \over 12}, \,\,\beta=
\frac{\sqrt{3} \alpha(\theta')^{1 \over 3}}{2^{4 \over 3}}
e^{-{5 i \pi \over 12}}. $$

From the other side we have (see (\ref{asAxt}, \ref{asBxt})): $$
{A \choose B} = { \chi_0 \sqrt{\varepsilon \over \theta} C_A +
\chi_1 \varepsilon^{3 \over 2} \sqrt{\theta} C_B \choose
\chi_0 \sqrt{\varepsilon \over \theta} C_B}
e^{i \left( x - {3 \pi \over 4} \right)} + $$
\begin{equation}
\label{asAB}
+{ \chi_0 \sqrt{\varepsilon \over \theta} C_A -
\chi_1 \varepsilon^{3 \over 2} \sqrt{\theta} C_B \choose
\chi_0 \sqrt{\varepsilon \over \theta} C_B}
e^{-i \left( x - {3 \pi \over 4} \right)},
\end{equation}
where $$\chi_0=\left( {2 \over \pi \sqrt{-\Gamma_0}} \right) ^{1
\over 2}, \,\,
\chi_1= \frac{\alpha(0)}{i \Gamma_0}
\left( {\sqrt{-\Gamma_0} \over 2 \pi} \right) ^ {1 \over 2}.$$

Our aim now is to express constants $C_A$, $C_B$ and $C_-$ through
$C_+$.  To find the corresponding expressions, one should match the
amplitudes of the incident and reflected waves in Eqs.(\ref{h}) and 
(\ref{asAB}) at a point $\theta_1$.  After some algebra, we have
\begin{equation}
\label{c-}
C_- = i C_+,
\end{equation}
\begin{equation}
\label{cb}
C_B = {\beta \theta_1^{1 \over 4} \over \chi_0 \varepsilon^{5
\over 2}} C_+,
\end{equation}
\begin{equation}
\label{ratio}
\frac{C_B}{C_A} = {i \chi_0 \over \varepsilon \theta_1 \chi_1}.
\end{equation}

The last step is to find the point $\theta_1$ at which the matching is
being made.  In solutions (\ref{h}) and (\ref{asAB}), the
phase-dependences are the same. However, phase-shifts appear in
the higher order approximations (see Eqs.(\ref{sigpm}) and (\ref{f}) for the
explicit expressions). We choose $\theta_1$ to be the point at which 
the phase-shifts for different asymptotics of the incident wave are the
same. It yields
\begin{equation}
\label{theta1}
\theta_1 = \frac{3 \varepsilon {\rm Im \,} f_{+} }
{ 2 \kappa \cos{\pi \over 12}}
\end{equation}

Mention, that our results are in agreement with all the
assumptions made in Sec. 4. From Eq.(\ref{ratio}), we see that $A
\sim \varepsilon^2 B$, from Eq.(\ref{asAB}), we obtain  
${\displaystyle dA \over \displaystyle
d\theta} \sim {\displaystyle A \over \displaystyle \varepsilon}$ and 
${\displaystyle dB \over \displaystyle d \theta}
\sim {\displaystyle B \over \displaystyle \varepsilon}$, 
as we have supposed. However, we
see that $\theta_1 \sim \varepsilon$ and this result is on
the border of applicability of asymptotics
(\ref{asH1}, \ref{asH2}). Strictly speaking, we should use the
exact form of Hankel functions to perform the matching, instead
of using their asymptotic representations. It could have some influence
on the values (\ref{cb}---\ref{theta1}), but not on (\ref{c-}).
For the case of simplicity, we avoid here the complification,
connected with the exact determination of $\theta_1$. However,
the physical results of our analysis remain applicable for
any values of $\theta_1$. Moreover, we emphasize, that
asymptotics (\ref{asH1}, \ref{asH2}) reflect the behavior
of Hankel function with a very good accuracy even for $|x| \sim 1$
(see Fig.~3) and hence, expressions (\ref{cb}--\ref{theta1}) are
well applicable as well.
                                                         
\section{Summary And Discussion}
This paper has derived the analytical expressions for the large-scale
magnetic field in the vicinity of the solar poles. These expressions
can be used to compare the relative magnitudes of the incident and
reflected waves.

Taking into account the fact that generalized momenta $k$ for the incident
and reflected waves differ only by sign at the pole [see
Eq.(\ref{kap})], we can estimate the ratio of the absolute values of
the amplitudes of the reflected and incident waves at 
point $\theta$ as follows (for the exact dependence see Fig.~3):
$$
R(\theta) \approx \exp{ \left(- {2 \over \varepsilon} \left| {\rm
        Im \,}k_0 \right| \theta \right)},$$
where for $k_0$ see
(\ref{kap}). Obviously, this ratio depends on $\theta$ and, as one can
see, the farther from the pole, the weaker the reflected wave
is compared to the incident wave.  Let us assume, that the
dynamo-number is $D=-10^4$ (or, equivalently, $\varepsilon\approx
0.05$) and the helicity coefficient is $\alpha(\theta)=\cos{\theta}$.
Evaluating $\theta_1$ explicitly we have $\theta_1 \approx
2.3^{\circ}$. This means that for $\theta < 2.3^{\circ} $ the solution
(\ref{as}) is applicable, for $\theta > 2.3^{\circ}$ one should use
expressions (\ref{fin}). Let $\theta=10^{\circ}$, than we obtain
$$ R(10^{\circ}) \approx 0.03.$$ It follows that the reflected
wave is about 30 times weaker than the incident wave. However,
the incident wave itself is rather weak near the pole against
the background of the dynamo wave in the main domain.  For
example, the incident wave at $\theta=10^{\circ}$ is approximately
10 times weaker than the dynamo wave in the domain of generation
(see Kuzanyan and Sokoloff, 1995). 
It follows, that the reflected wave is 300 times weaker than
the main wave.

Summarizing, we see that the solar magnetic field in the accepted
model is described with the aid of the three dynamo waves. The most
intensive wave propagates from high latitudes equator-wards. This wave,
of course, is well-known both from observations and from numerical
simulations. The second wave propagates in high latitudes pole-wards.
It has essentially lesser intensity than the first one. However, it
has been detected in the observations of the solar subpolar flares (see
Makarov and Sivaraman, 1983). In this paper we predict the existence
of another dynamo wave, which reflects from the pole and propagates
equator-wards.  We show that the reflected wave is very weak and it
decays exponentially propagating from the pole.  It is possible
however that the reflected wave could be detected in the future specialized
observations of the solar magnetic activity in the subpolar domain.

\begin{figure}
\epsfxsize=7cm
\centerline{\epsfbox{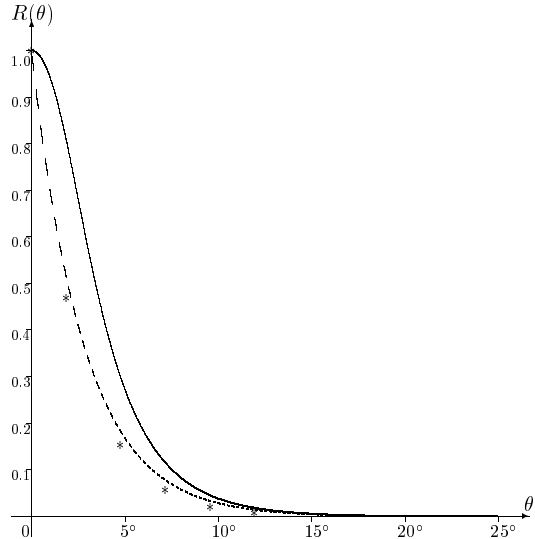}}
\caption{\rm\protect\footnotesize 
The ratio of magnitudes of the incident and reflected
waves as a function of latitude $\theta$. The solid line
represents dependence $R(\theta)$, obtained with use of 
(\protect\ref{fin}). The dashed line
corresponds to the approximate dependence $R(\theta)$,
calculated with the aid of asymptotics 
(\protect\ref{asH1}, \protect\ref{asH2}). Both curves
are extrapolated for large $\theta$.
Asterisks represnt some values $R(\theta)$ calculated using  the
WKB solution (\protect\ref{as}).  
The value $D = -10^4$ for all curves is accepted.}
\end{figure}

Using the solution obtained, we can also estimate the generation
threshold as a dynamo number, for which the generation of magnetic
field commences:
$$
\gamma_n \left( D^{\rm (cr)}_n \right) = 0.
$$
Hence, for the leading mode we have (n=0):
\begin{equation}
\label{Dcr}
\left| D^{\rm (cr)}_0 \right| \approx
- \left( \frac{ {\rm Re \,} \Gamma_1}{ {\rm Re \,} \Gamma_0}
\right) ^ 3 =
{2^{11 / 2} \over \alpha_{max}}.
\end{equation}
For $\alpha(\theta)=\cos(\theta)$ we have $\left| D^{\rm (cr)}_0
\right| \approx 90.5$.  Note, that this value is about two times
larger than the critical dynamo number for the Parker's equations.
Evaluating similar expressions $D^{\rm (cr)}_n$ for $n=1,2,3, \ldots$ and
taking into consideration existing estimates of $D$ for the Sun, one can 
conclude that only one additional mode (with $n=1$) is excited.
One should mention, that even provided that 
$\left| D^{\rm (cr)} \right| = 100$, we see,
that the true parameter of our asymptotic expansion is not very small  
and Eq.(\ref{Dcr}) should be
considered as a lower estimation for the critical dynamo number. 

As it was shown in Sec. 3, the value of $\Gamma_0$ depends on the
point $\theta'$ at which the matching of branches $k_3$ and $k_4$ is
made.  We have chosen this point to be the point of maximum of
$\hat\alpha(\theta)$.  The general case for the Parker's equations was
considered by Galitski and Sokoloff (1998). It was shown that the
corresponding eigensolutions can not be chosen to be smooth functions
for any other values of $\theta'$.  The same result can be obtained
for Eqs.(\ref{eqA}, \ref{eqB}). It means, that  operator $\hat H$ [see
(\ref{H})] does not possess any other discrete eigenvalues
except the ones obtained above (\ref{spectre}).

It is interesting also to compare our asymptotic theory with so-called
Maximally-Efficient Generation Approach (MEGA), see Ruzmaikin {\it et
  al}. (1990).  The latter suggests that the generated magnetic field is
localized in a small vicinity of the point at which the magnetic field
generation sources have maximum. Far from this point, the magnetic field
appears due to the diffusion mechanism only.  As one can see, the point of
maximum of our solution is shifted from the point of maximum of 
function $\hat\alpha(\theta)$ (this property of the solution was
emphasized by Kuzanyan and Sokoloff, 1995). It means, that in our case
the MEGA does not work quite well and one should be careful in application of
this approach.

Here, we considered dynamo equations in the kinematic approximation.
Let us note that in the recent paper of Meunier {\it et al.} (1997),
it was shown that in some nonlinear dynamo models, one can use the
kinematic approximation to describe the spatial profile of the
solution.  Certainly, in the nonlinear case, spectral parameter
$\gamma$ differs from eigenvalues (\ref{spectre}) (in the steady
nonlinear regime, $\gamma$ has only an imaginary part which determines the
period of the solar cycle).  However, our analysis of the dynamo wave
behavior near the pole remains applicable.

\vspace{1cm}
\noindent
{\it Acknowledgements}

{\footnotesize\noindent The authors are grateful to K.M.
Kuzanyan for helpful comments. This work was supported by
the Russian Foundation for Fundamental Research, grant No.
96-02-16252a}

$$$$ 
$$$$

\noindent
{\it References}

Bassom, A. P., Kuzanyan, K. M. and Soward, A. M., 1997: "A nonlinear
dynamo wave riding on a spatially varying background."
{\it Proc. R. Soc. Lond.} A, {\bf 455}, 1443---1481 (1999).

Brandenburg, A., "Solar Dynamos: Computational Background", in:
{\it Lectures on Solar and Planetary Dynamos}, 
(Ed. M. R. E. Proctor and A. D. Gilbert) pp. 117---159,
Cambridge University Press (1994).

Galitski, V. M. and Sokoloff, D. D., "Spectrum of Parker's
Equations", {\it Astron. Rep.} {\bf 42}, p. 127 (1998).

Jennings, R. L.,  "Symmetry breaking in a nonlinear $\alpha\omega$-dynamo",
{\it Geophys. Astrophys. Fluid Dyn.} {\bf 57}, 147---190 (1991).

Krause, F., R\"adler, K.-H., {\it Mean-Field Electrodynamics and
Dynamo Theory}, Oxford: Pergamon Press (1980).

Kuzanyan, K. M. and Sokoloff, D. D., "A dynamo wave in an inhomogeneous
medium", {\it Geophys. Astrophys. Fluid Dyn.} {\bf 81}, 113 ---
129 (1995).

Kuzanyan, K. M. and Sokoloff, D. D., "A dynamo wave in a
thin shell", {\it Astron. Rep.} {\bf 40}, pp. 425---430 (1996).

Landau, L. D. and Lifshitz, E. M., {\it Quantum Mecahnics},
Addison-Wesley, Mass. (1958).

Makarov, V. I. and Sivaraman, K. R. "Poleward migration of the
magnetic neutral line and reversals of the polar fields on the
Sun", {\it Sol. Phys.} {\bf 85}, 215---226 (1983).

Maslov, V. P. and Fedorjuk, M. V. {\it Semi-Classical Approximation
in Quantum Mechanics}, D. Reidel: Dordrecht (1981).

Meunier, N., Proctor, M. R. E., Sokoloff, D. D., Soward, A. M. and
Tobias, S. M., "Asymptotic properties of a nonlinear 
$\alpha\omega$-dynamo wave: period, amplitude and latitude
dependence." {\it Geophys. Astrophys. Fluid Dynam.} {\bf 86}, 249---285 
(1997).

Parker, E. N., Hydromagnetic dynamo models, {\it Astrophys. J.}
{\bf 122}, 293 --- 314, (1955).

Proctor, M. R. E. and Spiegel, E. A., "Waves of solar activity",
{\it In The Sun and Cool Stars: Activity, Magnetism, Dynamos}
(eds. I. Tuominen, D. Moss and G. R\"udiger), Lecture Notes in Physics.
Proceedings 380, pp. 117---128, Springer-Verlag (1991).

R\"udiger, G. and Brandenburg, A., "A solar dynamo in the
overshoot layer: cycle period and butterfly diagram", 
{\it Astron. Astrophys.} {\bf 296}, 557---566, (1995).

Ruzmaikin, A. A., Shukurov, A. M., Sokoloff, D. D. and
Starchenko, {\it Geophys. Astrophys. Fluid Dynam.} {\bf 52},
125---139,  (1990).

Stix, M., {\it The Sun: An Introduction}, Springer, Berlin (1989).

Tobias, S. M., "Properties of nonlinear dynamo waves", 
{\it Geophys. Astrophys. Fluid Dynam.} {\bf 86}, 287---343 (1997).

Tobias, S. M., Proctor, M. R. E. and Knobloch, E., "The role of absolute
instability in the solar dynamo", 
{\it Astron. Astrophys.} {\bf 318}, L55---58, (1997).

Worledge, D., Knobloch, E., Tobias, S. and Proctor, M. R. E.,
"Dynamo waves in semi-infinite and finite domains",
{\it Proc. R. Soc. Lond. A}, {\bf 453}, 119 --143 (1997).

Zeldovich, Ya. B., Ruzmaikin, A. A. and Sokoloff, D. D.,
{\it Magnetic Fields in Astrophysics}, Gordon and Breach, NY (1983).

\end{document}